\documentclass[aps,pra,twocolumn,amsmath,amssymb,superscriptaddress]{revtex4-2}

\usepackage{comment}

\usepackage{amsmath,amsfonts,amssymb}
\usepackage{physics}
\usepackage{graphicx}
\usepackage{setspace}
\usepackage{tocloft}
\usepackage[dvipsnames]{xcolor}

\usepackage{graphicx}
\usepackage{subcaption} 
\usepackage{dcolumn}
\usepackage[mathlines]{lineno}
\usepackage{physics}
\usepackage{hyperref}
\usepackage{multirow}
\usepackage{capt-of} 
\usepackage{bm}
\usepackage{epstopdf}
\usepackage{epsfig}
\usepackage{bbold}
\usepackage[normalem]{ulem}
\usepackage{caption}

\usepackage[dvipsnames]{xcolor}

\usepackage{amsmath}

\cftpagenumbersoff{figure}
\cftpagenumbersoff{table} 

\usepackage{CJK}
\usepackage{mathtools}
\usepackage[normalem]{ulem}
\usepackage{dcolumn}
\usepackage[mathlines]{lineno}
\usepackage{hyperref}
\usepackage{bm}
\usepackage{amssymb}
\usepackage{amsmath}
\usepackage{braket}
\usepackage{color}
\usepackage{xcolor}
\usepackage{soul} 
\usepackage{graphics}
\usepackage{bbm}

\begin{document} 
\title{Topological robustness of classical and quantum optical skyrmions in atmospheric turbulence}

\author{Zhenyu Guo}
\thanks{contributed equally to the work}
\affiliation{Centre for Disruptive Photonic Technologies, School of Physical and Mathematical Sciences, Nanyang Technological University, Singapore 637371, Singapore}

\author{Cade Peters}
\thanks{contributed equally to the work}
\affiliation{School of Physics, University of the Witwatersrand, Private Bag 3, Wits 2050, South Africa}

\author{Nilo Mata-Cervera}
\affiliation{Centre for Disruptive Photonic Technologies, School of Physical and Mathematical Sciences, Nanyang Technological University, Singapore 637371, Singapore}

\author{Anton Vetlugin}
\affiliation{Centre for Disruptive Photonic Technologies, School of Physical and Mathematical Sciences, Nanyang Technological University, Singapore 637371, Singapore}
\author{Ruixiang Guo}
\affiliation{Centre for Disruptive Photonic Technologies, School of Physical and Mathematical Sciences, Nanyang Technological University, Singapore 637371, Singapore}

\author{Pei Zhang}
\email[email:]{ zhangpei@mail.ustc.edu.cn}
\affiliation{Ministry of Education Key Laboratory for Nonequilibrium Synthesis and Modulation of Condensed Matter Shaanxi Province Key Laboratory of Quantum Information and Quantum Optoelectronic Devices, School of Physics, Xi’an Jiaotong University,
Xi’an 710049, China}

\author{Andrew Forbes}
\email[email:]{ andrew.forbes@wits.ac.za}
\affiliation{School of Physics, University of the Witwatersrand, Private Bag 3, Wits 2050, South Africa}
\email{andrew.forbes@wits.ac.za}

\author{Yijie Shen}\email{yijie.shen@ntu.edu.sg}
\affiliation{Centre for Disruptive Photonic Technologies, School of Physical and Mathematical Sciences, Nanyang Technological University, Singapore 637371, Singapore}
\affiliation{School of Electrical and Electronic Engineering, Nanyang Technological University, Singapore 639798, Singapore}

\date{\today}

\begin{abstract}
\noindent \textbf{The degradation of classical and quantum structured light induced by complex media constitutes a critical barrier to its practical implementation in a range of applications, from communication and energy transport to imaging and sensing.  Atmospheric turbulence is an exemplary case due to its complex phase structure and dynamic variations, driving the need to find invariances in light.  Here we construct classical and quantum optical skyrmions and pass them through experimentally simulated atmospheric turbulence, revealing the embedded topological resilience of their structure.  In the quantum realm, we show that while skyrmions undergo diminished entanglement, their topological characteristics maintain stable.  This is paralleled classically, where the vectorial structure is scrambled by the medium yet the skyrmion remains stable by virtue of its intrinsic topological protection mechanism.  Our experimental results are supported by rigorous analytical and numerical modelling, validating that the quantum-classical equivalence of the topological behaviour is due to the non-separability of the states and one-sided nature of the channel. Our work blurs the classical-quantum divide in the context of topology and opens a new path to information resilience in noisy channels, such as terrestrial and satellite-to-ground communication networks.
} 
\end{abstract}
 
\maketitle


\noindent 
Skyrmions, as topologically protected quasiparticles, have persistently driven research fervor from original particle physics~\cite{skyrmion1,sk2}, to condensed matter physics~\cite{sk3,sk4,sk5}, and photonics~\cite{gsk2}. Typically, magnetic skyrmions~\cite{msk1,msk2,msk3,msk4}, with their current-driven nanoscale solid-state manipulability, have revealed revolutionary application prospects in developing silicon-based high-density data memory devices~\cite{msk5,msk6,msk7}. In contrast to these typical magnetic skyrmions, with recent advances of structured light, skyrmions can be constructed in free-space propagating light waves~\cite{gsk3}, for both classical and quantum, opening opportunities for topologically enhanced large-scalar information transfer.
However, the topological resilience mechanism under atmospheric turbulence remains a critical scientific challenge to be addressed.

By leveraging the unique degrees of freedom (DoF) intrinsic to photonic systems, researchers have successfully achieved both controlled generation and dynamic manipulation of skyrmions across multidimensional parameter spaces, encompassing spin angular momentum \cite{sksam,sksam2,sksam3,wu2025photonic}, Stokes parameters \cite{skst1,skst2,skst3,skst4,skst5}, electric or magnetic field vectors \cite{shen2021supertoroidal,liu2022disorder,wang2024observation}, energy flows~\cite{wang2024topological,chen2025topological}, and momentum-space pseudospin vectors \cite{sksk1,sksk2,sksk3,rao2025meron}.
While previous studies have consistently conceptualized skyrmions as localized fields or particles, recent groundbreaking work has experimentally demonstrated quantum skyrmions by harnessing entanglement between orbital angular momentum (OAM) and polarization in biphoton systems \cite{quansk1}.
The innovation lies in the global encoding of topological information through entangled correlations in the spatial and polarization degrees of freedom within biphoton states, representing a fundamental departure from conventional single-particle parameter-space distributions, and allowing protection of quantum information against noise \cite{ornelas2025topological}.

Here we transmit quantum skyrmions through experimentally simulated atmospheric turbulence and demonstrate their robustness in realistic environments. 
Our experiment is theoretically underpinned by a correlation framework that connects perturbations in complex media to variations in quantum states via spiral imaging theory, while coordinate transformation analysis systematically elucidates the fundamental principles governing topological robustness. 
In parallel, we create a local equivalent of the quantum state, a classical optical skyrmion, and pass it through the same channel, again showing robustness.  We highlight that both cases can be modeled as one-sided channels, leading to equivalent behaviour.  Our research provides a comprehensive validation through an integrated theoretical and experimental approach of the topological robustness of skyrmions against complex media such as atmospheric turbulence.

\section*{Results}

\noindent \textbf{Classical-quantum equivalence of topologies.}  Spin-textured structured light can be written compactly as $\ket{\Psi} =  \ket{M_1}_A \ket{e_1}_B  + \ket{M_2}_A \ket{e_2}_B$, where the polarisation DoF is expressed as any pair of orthonormal states, $\{ \ket{e_1}$ and $\ket{e_2} \}$ while the spatial mode DoF is given by the orthonormal basis states $\{ \ket{M_1}$ and $\ket{M_2} \}$.  As a quantum state the subscripts refer not only to two DoFs but also two photons, $A$ and $B$, and so the non-separability is non-local, while classically the non-separability is local \cite{Spreeuw1998,forbes2019classically,shen2022nonseparable}.  

It is possible to imbue such states with a topology, local in the classical case and non-local in the quantum case.  Without loss of generality we will henceforth do so by considering the state

\begin{equation}
	\ket{\Psi} =  \lambda_1 \ket{\ell_1}_A \ket{H}_B  + \lambda_2 \ket{\ell_2}_A \ket{V}_B \, ,
	\label{eq1}
\end{equation}
where $\lambda_i$ represents the normalized complex coefficients, $\ket{H}$ and $\ket{V}$ correspond to mutually orthogonal horizontal and vertical polarization states, and  $\ket{\ell} \equiv \text{LG}_\ell(r,\phi)$ for classical light and 
$\ket{\ell} \equiv \int |\text{LG}_\ell (r,\phi)| \exp^{i \ell \phi} \ket{r,\phi} rdrd\phi$ for quantum states.

When $|\ell_1|\neq|\ell_2|$, these structures form a mapping from the transverse plane $\mathcal{R}^2$ to the Poincar\'{e} sphere $\mathcal{S}^2$, defining a skyrmionic topology. The topological invariant $N$, referred to as the skyrmion number or wrapping number, uniquely characterizes each topology and quantifies how many times one wraps around $\mathcal{S}^2$ after completely traversing $\mathcal{R}^2$ through a stereographic projection. The wrapping number can be calculated using,
\begin{equation}
    N = \frac{1}{4\pi} \int_{\mathcal{R}^2} \epsilon_{ijk} S_i \frac{\partial S_j}{\partial x} \frac{\partial S_k}{\partial y} \text{d}x \text{d}y \,,
    \label{eq:Skyrme wrapping}
\end{equation}
where $\epsilon_{ijk}$ is the Levi-Civita symbol and $S_i$ with $i=1,2,3$ are the locally normalized Stokes parameters such that $\Sigma_{i=1}^3 S_i = 1$. This normalization ensures that the mapping onto $\mathcal{S}^2$ always maps onto a unit sphere. For the above states, equation \ref{eq:Skyrme wrapping}  simplifies to: (See method for detailed derivation)
\begin{equation}
    N = n |\ell_1-\ell_2| \,,
\end{equation}
where $n = -1$ if $l_2 > l_1$ and $n=1$ if $l_2 < l_1$. 
\begin{figure}[htbp]
\centering\includegraphics[width=\linewidth]{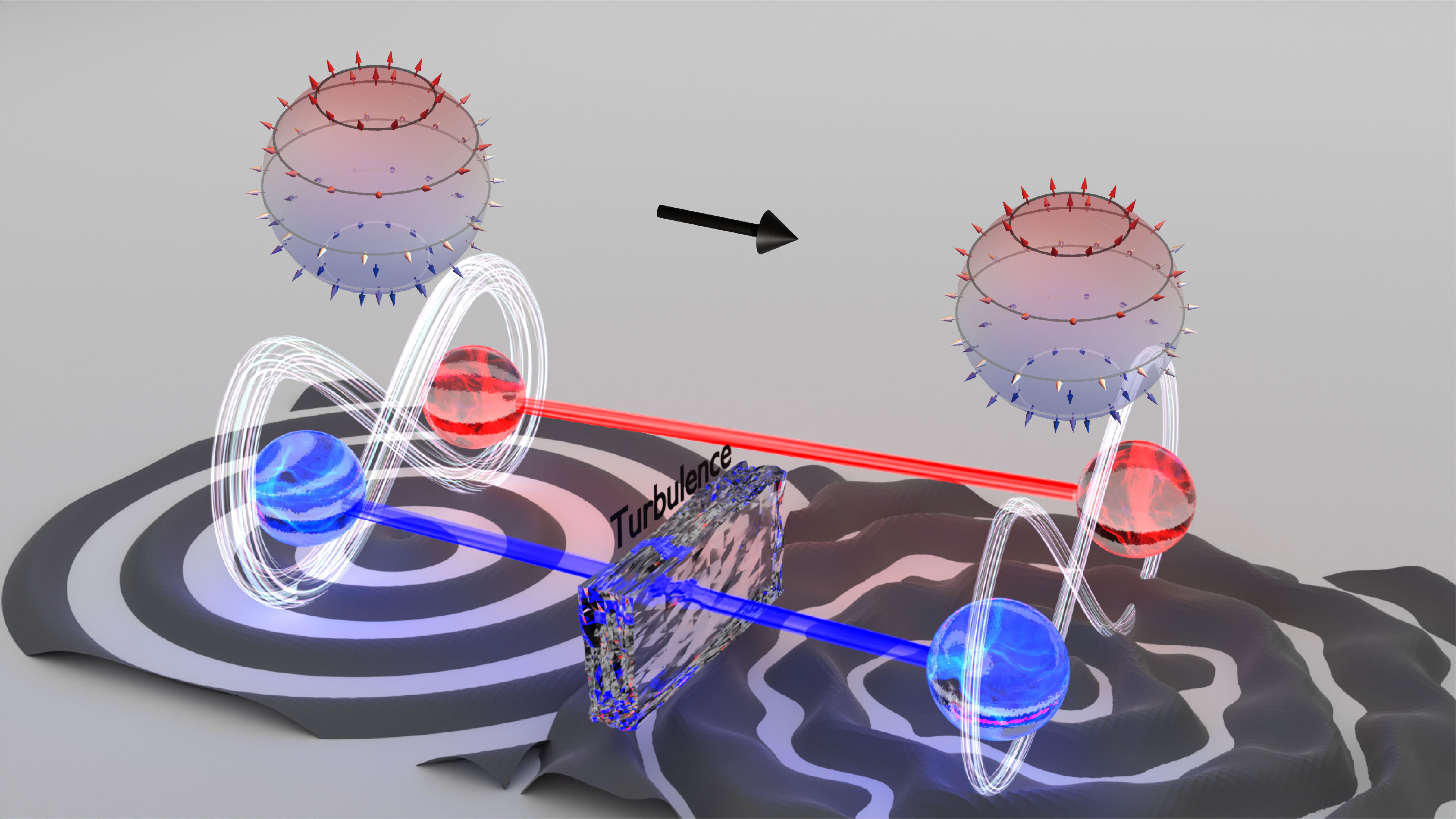}
\caption{\textbf{Schematic of skyrmions in turbulence.} In quantum skyrmion systems, the blue and red spheres denote photon A carrying OAM and photon B encoding polarization information, respectively. In their classical counterparts, these spheres represent the beam’s OAM and polarization degrees of freedom. Strikingly, both systems exhibit the same topological invariance: although turbulence can degrade quantum entanglement or classical correlations, the skyrmion number remains strictly conserved.}
	\label{figure1}
\end{figure}

Now imagine that our state is passed through a one-sided channel, where one DoF is affected and the other not.  Atmospheric turbulence is such a channel since the spatial mode (DoF $A$) is distorted, while the polarisation (DoF $B$) is not.  We can frame a general equivalence to both the classical and quantum states and use this to provide an agnostic analysis, where pure topological quantum states are equated to their classical counterparts by virtue of non-separability \cite{Ndagano2017}. 
As shown in Figure \ref{figure1}, under turbulent conditions, both quantum and classical skyrmions exhibit remarkable robustness in their topological properties, with the topological charge remaining invariant.
Any one-sided channel is unitary to any input pure state since it may be written as a positive trace-preserving map, ensuring that the output must also be a pure state.  The Choi-Jamiolkowski isomorphism \cite{jiang2013channel} establishes a correspondence between the channel operator, $T_A$, and a quantum state, so that a measurement on one returns the other.  The state after the channel, both classical and quantum, is then $\ket{\Psi_\text{out}} = (T_A\otimes \mathbb{1}_B)\ket{\Psi_\text{in}}$, where $\mathbb{1}_B$ is the identity operator for DoF $B$ with the subscript in $\ket{\Psi}$ indicating the input and output states.  This shows that the spin texture of the field will in general change, but that the distorted spatial modes at the output will still be orthogonal, thus maintaining the non-separability (inhomogenous nature of the spin texture) of the state \cite{nape2022revealing}. The open question which we seek to address here is whether this implies that the topology, as measured by $N$, changes or remains intact? 

\vspace{0.5cm}

\section*{Quantum Skyrmions in Turbulence}

\noindent \textbf{Topological resilience of quantum skyrmions to turbulence.}  
\begin{figure*}[htbp]
\centering\includegraphics[width=0.83\linewidth]{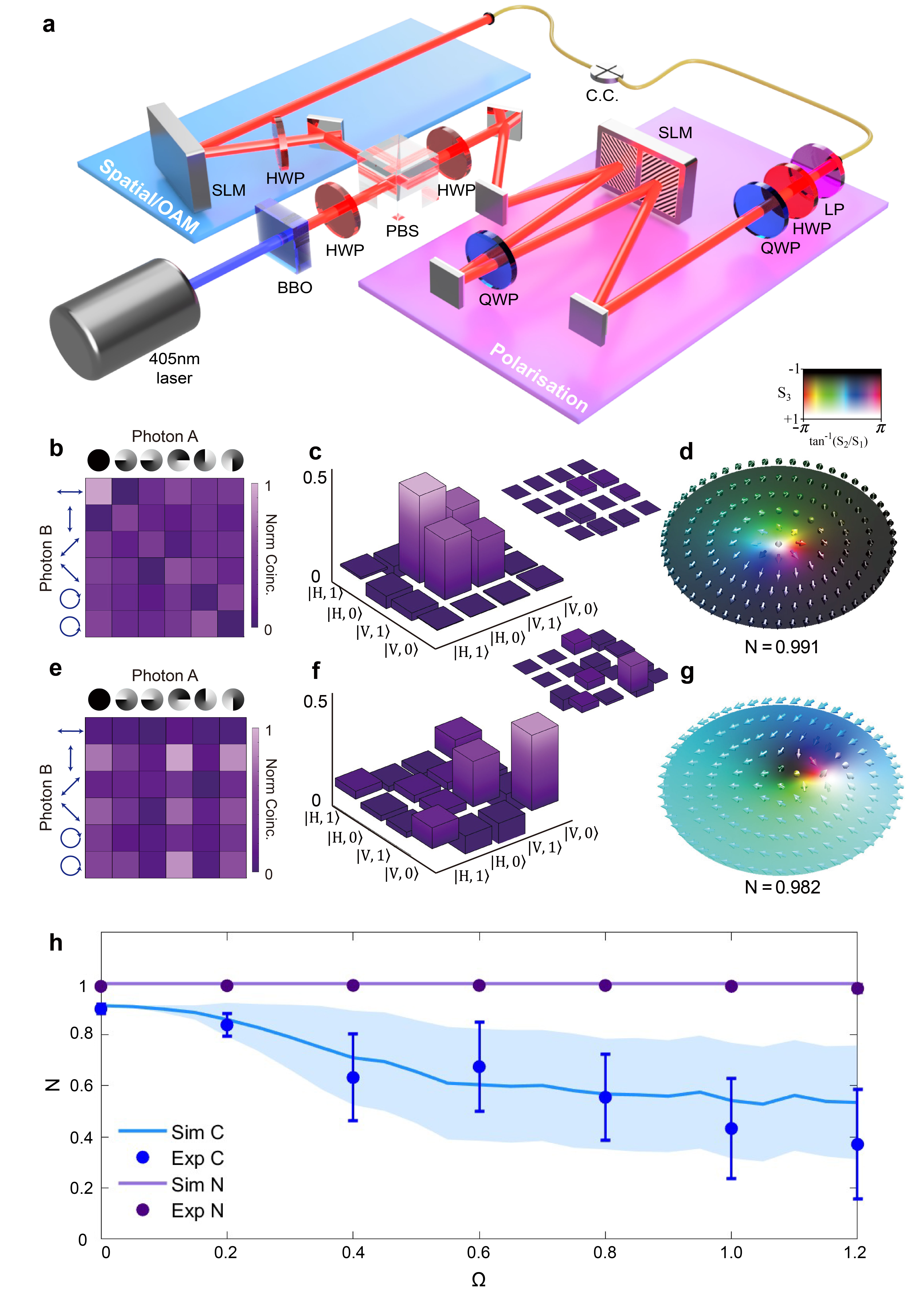}
\caption{\textbf{Quantum Experimental Results.}
\textbf{a} Experimental setup. BBO, $\beta$-barium borate crystal; QWP, quarter-wave plate; HWP, half-wave plate; PBS, polarizing beam splitter; SLM, spatial light modulator; LP, linear polarizer.
\textbf{b}-\textbf{d} present the joint measurement data obtained from quantum state tomography of the skyrmion $|\Psi\rangle = \tfrac{1}{\sqrt{2}} ( |0\rangle_A |H\rangle_B + |1\rangle_A |V\rangle_B )$, the reconstructed density matrices, and the corresponding spin-textured fields in the absence of turbulence.
\textbf{e}-\textbf{g} Experimental tomography results, reconstructed density matrices, and corresponding spin-textured fields under turbulence strength $2w/r_0 = 1.2$.
\textbf{h} Evolution of skyrmion number $N$ and concurrence $C$ versus turbulence strength $\Omega = 2w/r_0$. The solid curve denotes theoretical predictions, and the blue shaded region indicates the error range from 200 independent simulations.}
\label{figurefinal}
\end{figure*}
We first answer this question in the context of quantum light, using recently developed formulation \cite{de2025quantum}. Photons, as flying qubits, play a pivotal role as information carriers in various quantum information tasks. However, environmental disturbances such as atmospheric turbulence significantly degrade their coherence, posing a major obstacle to the construction and advancement of practical quantum networks.
To address this challenge, extensive efforts have been devoted by researchers, including the introduction of eigenmodes of turbulence \cite{apap1,peters2025tailoring}, demonstrations of the disturbance-resistant advantages of vector structured light \cite{nape2022revealing}. 
Nevertheless, an effective strategy to directly mitigate turbulence effects on quantum states remains elusive. The advent of quantum skyrmions, however, offers new opportunities for achieving robust quantum state transmission. The hybrid entanglement intrinsic to quantum skyrmions naturally imparts photons carrying polarization DoF with enhanced resilience against turbulence. In contrast, photons encoded solely with OAM remain highly susceptible to perturbations, resulting in significant degradation of quantum entanglement. Remarkably, owing to the global nature of the mapping encoded in the topological features of skyrmions, the associated topological charge exhibits unexpected robustness, highlighting its strong potential for reliable state transmission in turbulent environments.
After being subjected to turbulence, the initial quantum state experiences significant perturbations. This turbulence-induced evolution can be effectively described using the theory of spiral imaging as follows \cite{spiralima}:
\begin{equation}
\left|l\right\rangle \rightarrow \sum_{l^{\prime}} \alpha_{l^{\prime}}\left|l^{\prime}\right\rangle,
\end{equation}
where $\alpha_{l^{\prime}}$ denotes the complex amplitude coefficient. This process indicates that turbulence causes the initial OAM eigenstate to disperse into other eigenstates. Since we consider only qubits, the system's quantum state can be expressed as:
\begin{equation}
\begin{aligned}
|\Psi\rangle & =\lambda_1\left|l_1\right\rangle_A|H\rangle_B+\lambda_2\left|l_1\right\rangle_A|V\rangle_B \\
& +\lambda_3\left|l_2\right\rangle_A|H\rangle_B+\lambda_4\left|l_2\right\rangle_A|V\rangle_B,
\end{aligned}
\end{equation}
where $\lambda_i$ represents the normalized complex coefficients.
Direct observation shows that the turbulence-induced perturbation of the quantum state manifests primarily as the introduction of additional spatial modes in photon A.
The resulting increase in state-space complexity leads to a marked degradation of the system’s entanglement.

The topological properties of a quantum system are encoded in the global features of its wave function, rendering its topological characteristics insensitive to local perturbations. This invariance is exemplified by the skyrmion number, a topological invariant that remains unchanged under coordinate transformations. Mathematically, it can be expressed as:
\begin{equation}
\begin{aligned}
N & =\frac{1}{4\pi} \int_{\mathcal{R}^2} \epsilon_{ijk} S_i \frac{\partial S_j}{\partial x} \frac{\partial S_k}{\partial y} \text{d}x \text{d}y \\
& =\frac{1}{4\pi} \int_{\mathcal{R}^2} \epsilon_{ijk} S_i \frac{\partial S_j}{\partial x^{\prime}} \frac{\partial S_k}{\partial y^{\prime}} \text{d}x^{\prime} \text{d}y^{\prime}.
\end{aligned}
\label{eq11}
\end{equation}
Therefore, by characterizing the effect of turbulence on the quantum state through coordinate transformations (as detailed in the supplementary information), one can rigorously demonstrate that quantum skyrmions retain their topological robustness under turbulence, owing to the invariance of the skyrmion number under such transformations.


\vspace{0.5cm}

\noindent \textbf{Experimental demonstration.}  The experimental procedure consists of three stages: the preparation of quantum skyrmions, the introduction of turbulence, and the measurement of quantum states. The experimental setup is depicted in Figure \ref{figurefinal} \textbf{a}. Specifically, a 405 nm continuous-wave laser serves as the pump source, and spatial filtering via a single-mode fiber is employed to produce a Gaussian spatial mode. This configuration allows precise control over the spiral spectrum bandwidth of the entangled photon pairs, thereby enabling direct manipulation of the skyrmions’ topological charge.
The filtered pump beam is then directed onto a 3‑mm-thick $\beta$-barium borate nonlinear crystal, where type-I spontaneous parametric down-conversion generates entangled photon pairs exhibiting OAM correlations.
After passing through the band-pass filter, only the 810 nm parametric photons are retained, and the quantum state at this stage can be expressed as:
\begin{equation}
|\Psi\rangle=\sum_{l=-\infty}^{\infty} c_l\left|l\right\rangle_A\left|-l\right\rangle_B.
\end{equation}

To realize the preparation of quantum skyrmions, the initial OAM-entangled state must be transformed into a hybrid-entangled state. In the experiment, a beam splitter spatially separates the photon pairs: photon A retains its OAM DoF to carry spatial information, while photon B undergoes a fully digital spatial–polarization conversion process, coupling its OAM DoF to a specific polarization state.
The conversion process implemented by a spatial light modulator (SLM) is based on its polarization-selective operating characteristics\cite{spc}. 
The implementation proceeds as follows: first, horizontally polarized incident photons undergo initial phase modulation by the SLM. Upon reflection, the remaining vertical polarization component is converted into a modulable state via an optical system comprising a quarter-wave plate (QWP) and a mirror, thereby enabling complete control over the orthogonal polarization states.
Upon completion of the conversion process, the quantum skyrmion was experimentally demonstrated.

Photons carrying OAM first pass through an SLM followed by a coupling-lens system before being coupled into a single-mode fiber. Notably, the SLM is configured with both holographic gratings for quantum state tomography and additional phase-modulation gratings designed to simulate atmospheric turbulence (see supplementary information). Meanwhile, polarization-encoded photons are directed through a Stokes parameter measurement system comprising a QWP, HWP, and linear polarizer, after which they are coupled into another fiber via a lens assembly. Experimental measurements are obtained using coincidence counting techniques based on detection signals from single-photon detectors positioned at both ends of the setup.

Quantum state tomography was performed by conducting mutually unbiased basis measurements on both photons’ OAM and polarization DoF (see supplementary information). 
First, we performed quantum state tomography on quantum skyrmion state $|\Psi\rangle = \tfrac{1}{\sqrt{2}} ( |0\rangle_A |H\rangle_B + |1\rangle_A |V\rangle_B )$ in the absence of turbulence, with the joint measurement data shown in Figure \ref{figurefinal} \textbf{b}.
Based on this dataset, the quantum state's density matrix was reconstructed using maximum likelihood estimation, as shown in Figure \ref{figurefinal} \textbf{c}. Analysis of the density matrix further allowed us to extract the spin texture characteristics of the quantum skyrmion. Notably, under turbulence-free conditions, the measured spin texture (Figure \ref{figurefinal} \textbf{d}) exhibits a clearly defined Néel-type topological structure.
Subsequently, we performed state tomography on quantum skyrmions under varying turbulence intensities to investigate their topological and entanglement properties. Figures \ref{figurefinal} \textbf{e}–\textbf{g} show the experimentally obtained state tomography data, reconstructed density matrices, and corresponding spin-textured fields under turbulence strength $2w/r_0 = 1.2$. Although turbulence induces significant perturbations to both the quantum state and the spin texture, the topological charge across the entire field remains conserved.
In the experiment, we quantified the entanglement strength of the two-qubit system using concurrence, which ranges from 0 to 1, with 0 indicating no entanglement and 1 indicating maximal entanglement. As shown in Figure \ref{figurefinal} \textbf{h}, even as increasing turbulence deforms the quantum texture and reduces the concurrence, the topological charge of the quantum skyrmion remains rigorously preserved. This robust topological protection has been confirmed through both experiment and simulation, following the mechanism described in equation (\ref{eq11}): nonlocal correlations transform turbulence-induced perturbations into smooth coordinate transformations, thereby ensuring strict conservation of topological properties.


\section*{Classical Skyrmions in Turbulence}


\begin{figure*}[htpb]
\centering\includegraphics[width=\linewidth]{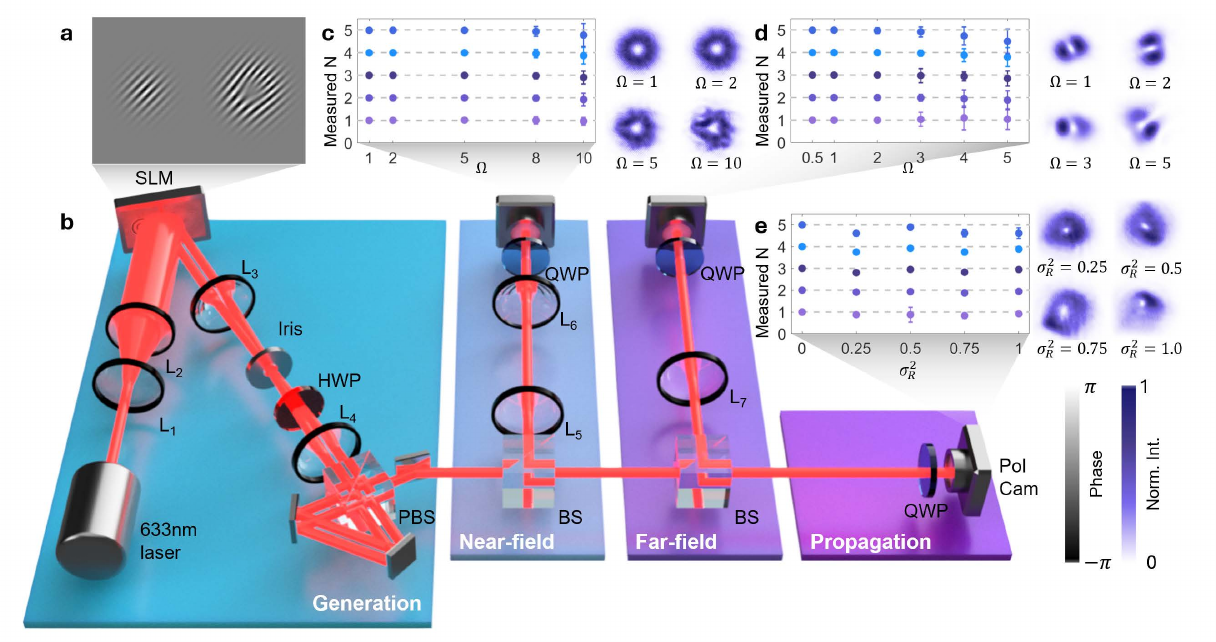}
\caption{\textbf{Classical experimental results.} \textbf{a} Complex amplitude hologram used to generate the vector beam $\ket{\Psi} = \left( \ket{LG_0}\ket{V} + \ket{LG_2}\ket{H} \right) / \sqrt{2}$. \textbf{b} Experimental setup showing the generation of two scalar beams that are vectorially combined using a Sagnac interferometer. \textbf{c} Experimental results showing the mean measured skyrmion number of beams in near-field turbulence of varying turbulence strength $\Omega = 2w/r_0$. \textbf{d} Experimental results showing the mean measured skyrmion number of beams in far-field turbulence of varying strength $\Omega$. \textbf{e} Experimental results showing the mean measured skyrmion number of beams after propagating through simulated turbulence channel of 1~m length and varying Rytov variance $\sigma_R^2$.}

    \label{fig:Classical results}
\end{figure*}

\vspace{0.5cm}
\begin{figure*}[tpb]
\centering\includegraphics[width=0.9\linewidth]{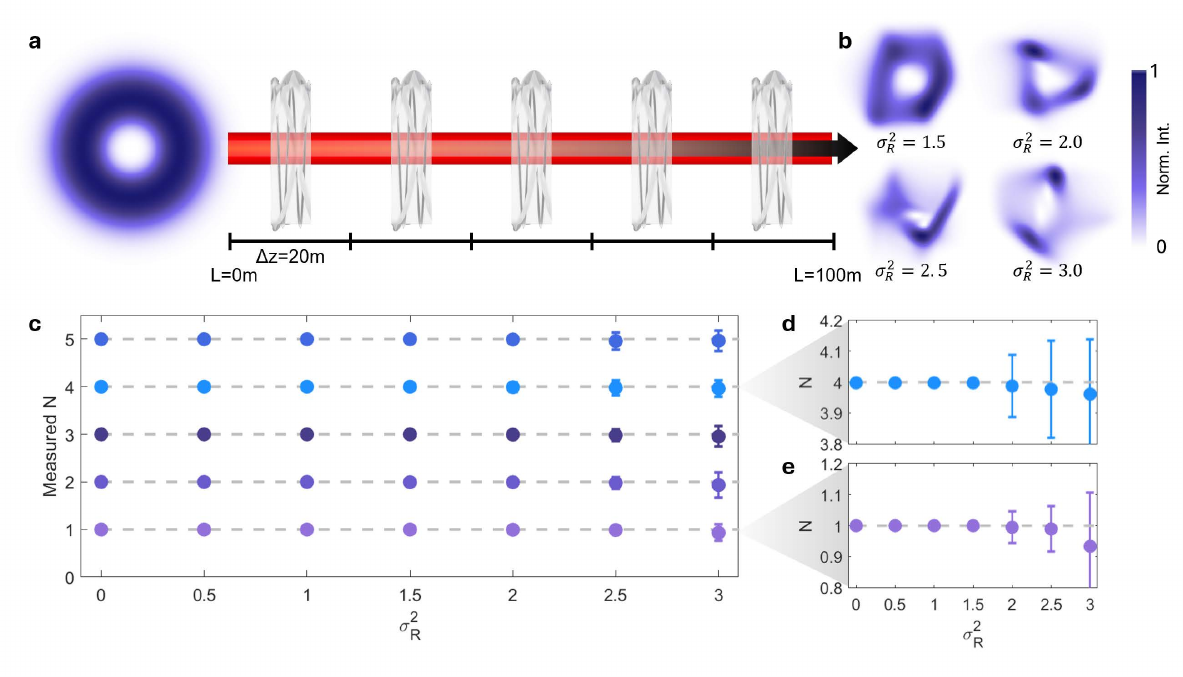}
	\caption{\textbf{Skyrmions through thick turbulence.} \textbf{a} Sketch of the geometry of the wave-optics simulations used to simulate thick medium turbulence. The total channel length was 100~m, with 5 turbulence phase screens evenly spaced along the path, separated by 20~m. \textbf{b} Example simulated intensity distributions of and $LG_2$ beam after propagating through turbulent channel of varying strengths. \textbf{c} Results showing the mean measured skyrmion number of beams after propagating through a numerically simulated thick-turbulence channel of 100~m length and varying Rytov variance $\sigma_R^2$.  \textbf{d} Zoomed in plots of the simulated results for $N=4$. \textbf{e} Zoomed in plots of the simulated results for $N=1$. }
    \label{fig:MPS results}
\end{figure*} 
\noindent \textbf{Robustness of classical skyrmions in atmospheric turbulence.} Investigating optical skyrmions in the classical regime allows us to easily asses their robustness in a basis independent manner. In the quantum regime, spatial projective measurements are made only onto the OAM subspace of choice. In the classical regime, we can make use of polarisation optics and a camera to directly obtain the spatially resolved Stokes parameters. In this way, we may observe the effects of inter-modal coupling across all basis modes with four simple intensity measurements, and not just coupling within a specific subspace. This provides a complementary analysis of the topological robustness, as different scenarios and applications will require and may benefit from different measurement approaches. The close analogy between classical vector beams and quantum hybrid entangled states implies that the results in this article are directly applicable and easily adaptable from one regime to another \cite{shen2022nonseparable,peters2023spatially}.

Classical optical Skyrmions were generated using a SLM holograms with a complex amplitude modulation scheme \cite{arrizon2007pixelated} in combination with a Sagnac interferometer. An example SLM hologram is shown in Figure \ref{fig:Classical results} \textbf{a} and the experimental setup is shown \ref{fig:Classical results} \textbf{b} with further details provided in the Methods section. Figure \ref{fig:Classical results} \textbf{c} shows a summary of results in the near-field (i.e. in the image plane of the SLM) where each data-point represents the mean measured $N$ over 100 phase screen realizations and the error bars denote the standard deviation. In this regime, we expect the added turbulent phase screen to act only as a phase perturbation and thus have no effect on the skyrmion number \cite{peters2025seeing}. We tested 5 distinct skyrmion topologies $N = 1$ to $N=5$ across 5 different turbulence strengths. The turbulence strengths were quantified by the unitless parameters $\Omega = 2w/r_0$, where $w$ is the second moment radius of the total beam and $r_0$ is the Fried parameter of the phase screen. We see over the range tested, $N$ remains almost perfectly unchanged with only slight deviations evident at $\Omega = 10$ where the correlation length of the medium is an order of magnitude smaller than the transverse beam size and constitutes a very strong perturbation. Figure \ref{fig:Classical results} \textbf{d} shows a summary of results in the far-field where each data-point represents the mean measured $N$ over 100 phase screen realisations and the error bars denote the standard deviation. We see that for low and moderate strengths $\Omega = 0.5,1$ and $2$, the skyrmion number remains almost perfectly unchanged for all 5 of the tested topologies. As the strength increases past these, we see that there is some noticeable variation in the skyrmion number up until $\Omega = 5$, where the error bars begin to overlap, indicating a regime where one can no longer confidently identify the original topological number. The far-field represents the extreme limit of propagation. In order to investigate a regime that lies between the near field and the far-field, we also tested the robustness of optical skyrmions through an experimentally simulated $100$~m channel using a 1~m propagation distance in the lab. These results were performed in such a manner such that they can be generalized to channels of arbitrary lengths but of same Rytov variance $\sigma_R^2$ according to Fresnel scaling procedure described in Ref \cite{peters2025structured}. The Rytov $\sigma_R^2$ variance is paramount in quantifying the turbulence strengths as it accounts not only for the strength of the phase perturbation $r_0$, but also scintillation due to propagation. Figure \ref{fig:Classical results} \textbf{e} shows a summary of these experimental results using only one phase screen, where each data-point represents the mean measured $N$ over 100 phase screen realizations and the error bars denote the standard deviation. We see that for all topologies except $N=5$, we see excellent agreement between the encoded and measured skyrmion number. The error bars are also barely noticeable, indicating almost negligible variation over multiple realizations and demonstrating remarkable stability of the topology. In Figure \ref{fig:Classical results} \textbf{d} and \textbf{e}, $N=5$ demonstrates a small but noticeable decay as the aberration strength increases. This can be attributed to the fact that higher order topologies exhibit polarisation structures that change rapidly across the transverse plane. When these structures are distorted and no longer symmetrical, it can be challenging to reliably identify all of the polarisation singularities needed to accurately determine the skyrmion number. This can possibly be improved by more intelligently choosing the component scalar modes to allow for a more spread out singularity distribution. Such state engineering approaches have already shown an increase in the reliability of measuring other topological structures such as optical knots \cite{pires2025stability}. 

\noindent \textbf{Robustness through extended turbulent media.} So far, we have only investigated the effect of turbulence as acting at single plane with the single phase screen approximation. However, to investigate the resilience of optical Skyrmions through thick/extended media such as long path atmospheric turbulence, one must make use of a multiple phase screen approach. Due to the inefficiencies of polarisation insensitive SLMs such as digital micro-mirror devices, we opted to make use of numerical wave-optics simulations, consisting of 5 phase screens evenly space out over a $L=100$~m channel as seen in Figure \ref{fig:MPS results} \textbf{a}. Following the approach outlined in Ref. \cite{peters2025structured}, we investigated turbulence channels of distortion strengths ranging from $\sigma_R^2 = 0.5$ to $\sigma_R^2=3.0$ with full simulation parameters and details in the supplementary information. We show example intensity distributions of a $LG_2$ beam after propagating through the simulated channel in Figure \ref{fig:MPS results} \textbf{b} for distortion strengths $\sigma_R^2 = 1.5$, $\sigma_R^2 = 2.0$, $\sigma_R^2 = 2.5$ and $\sigma_R^2 = 3.0$, demonstrating the significant distortion induced on the input beam at all strengths. We show the measured skyrmion number of the beams after propagating through the channel in Figure \ref{fig:MPS results} \textbf{c}, where each data point represents the average over 100 independent realizations of the turbulent channel and the error bars give the standard deviation. We see that over all turbulence strengths and for all encoded skyrmion numbers, the topology of the beam is extremely well maintained at the output of the channel, with very little deviation, even at considerably high distortion strengths. Figure \ref{fig:MPS results} \textbf{d} and \textbf{e} show zoomed in view of the results for encoded skyrmion number of $N=1$ and $N=4$ respectively. We see that the topology only begins to noticeably vary in both cases at a distortion strength of $\sigma_R^2=2$ and greater. However, the deviation is still significantly small when compared to the value of the skyrmion number and, as a consequence, the topologies are still easily distinguished from each other over the entire range of tested strengths.

The robustness of optical skyrmions in extended propagation through complex channels has seen very little study, primarily due to the complex dynamics of spatially varying polarization fields as they diffract. In general, it cannot be assumed that optical skyrmions will remain robust when propagating through through arbitrary complex media, even if that medium is unitary. A simple example of this would be a cylindrical mode converter where an input Laguerre-Gaussian beam is converted to a Hermite-Gaussian beam either at a plane \cite{vaity2013measuring} or permanently \cite{beijersbergen1993astigmatic}. In such simple optical systems, the mode conversion will lead to the a decay and eventual loss of the skyrmionic topology. It is therefore surprising to see that optical skyrmions are robust through atmospheric turbulence, which are often considered to be an extreme example of a complex channel. This is even more surprising when one considers that the Noll covariance matrix of atmospheric turbulence weights astigmatism (the key aberration in mode converters) as its most significant component aberration \cite{noll1976zernike,roddier1990atmospheric}. Mode conversion typically manifests only in the far-field \cite{buono2022eigenmodes}, explaining why the propagation results in Figure \ref{fig:Classical results} \textbf{e} and Figure \ref{fig:MPS results} \textbf{c} exhibit almost perfect robustness when compared to the far-field results in Figure \ref{fig:Classical results} \textbf{d}. Even so, we only see the measured $N$ show significant variation around the encoded value in strong conditions in far-field, where $\Omega = 5$ and only observe the mean measured deviating from the encoded value for higher order topologies $N=5$. These results then provide strong evidence for the robustness of optical skyrmions for a large range of turbulent conditions and for vast majority of real-world scenarios.\\

\section*{Discussion and Conclusion}
Our experimental results demonstrate that the entanglement characteristics of quantum skyrmions inevitably degrade during turbulent propagation—a manifestation of the intrinsic fragility of quantum states that poses significant challenges for entanglement-dependent quantum information tasks. Crucially, when topological information is encoded in the global mapping of biphoton systems, the topological charge exhibits remarkable robustness even under strong turbulence conditions. Parallel studies reveal that while the spatial modes of classical skyrmions undergo turbulence-strength-dependent distortion, their topological charge—constructed through Stokes vector-spatial mode mapping—remains equally preserved.
The establishment of a universal theoretical framework represents a key breakthrough: by leveraging the essential feature of quantum state non-separability, we achieve fundamental equivalence between pure topological quantum states and their classical counterparts. This discovery unveils a profound shared characteristic of quantum and classical skyrmions—the unitary nature of any single-sided quantum channel acting on input pure states. This unitarity originates from the channel's representation as a trace-preserving positive map, thereby guaranteeing the output state's preservation of purity. 

This study holds profound significance in multiple dimensions: Firstly, we have for the first time systematically elucidated the fundamental connection between quantum/classical skyrmion systems and their topological structures, a breakthrough that not only deepens the theoretical foundation of quantum-classical correspondence but also establishes a novel paradigm for unified description of both systems from the perspective of topological invariance. Secondly, we have innovatively proposed a new type of turbulence-resistant information carrier, whose unique topological protection mechanism opens up new developmental pathways for both classical optical communication and quantum information technologies. This groundbreaking advancement is expected to drive revolutionary innovations in several key areas: highly robust free-space optical communication systems, interference-resistant quantum key distribution networks.
More importantly, the theoretical framework we have established provides a solid foundation for developing next-generation noise-resistant quantum information processing technologies, with potential applications including but not limited to: quantum sensing in complex environments, noise-tolerant quantum computing, and all-weather quantum communication systems. These innovative achievements will significantly accelerate the practical implementation of both classical and quantum technologies under real-world conditions. The topological protection mechanism we discovered demonstrates remarkable universality, maintaining its protective effects across different physical platforms and environmental conditions, which may inspire new research directions in topological photonics and quantum information science.

\vspace{0.5cm}
\newpage
\section*{Materials and correspondence}
Correspondence and requests for materials should be addressed to AF and YS.


\section*{Acknowledgments}
AF thanks the NRF-CSIR Rental Pool Programme. YS thanks the Nanyang Assistant Professorship Start Up Grant, Singapore Ministry of Education (MOE) AcRF Tier 1 grants (RG157/23 \& RT11/23), and Singapore Agency for Science, Technology and Research (A*STAR) MTC Individual Research Grants (M24N7c0080).

\section*{Authors' contribution}
ZG performed the quantum experiments. CP performed the classical experiments. All authors contributed to data analysis, assisting experimental setup and writing of the manuscript. AF and YS conceived the initial idea and supervised the project.

\section*{Competing financial interests}
The authors declare no financial interests.

\newpage

\section*{Methods}
\subsection{Theory of Quantum Skyrmions.} 
Quantum skyrmions are generated through a nonlocal mapping between spatial coordinates and Stokes parameters, mediated by the entanglement of OAM and polarization in biphoton systems, resulting in topologically entities.
The eigenstates of OAM can be expanded in the position space using Laguerre-Gaussian functions, expressed as $\left|L G_l\right\rangle=\int\left|L G_l\left(\overrightarrow{r_A}\right)\right| e^{i l \phi}\left|\overrightarrow{r_A}\right\rangle d^2 r_A$.
The quantum skyrmion state can be expressed as\cite{quansk1}:
\begin{equation}
	|\Psi\rangle=\lambda_1\left|L G_{l_1}\right\rangle_A|H\rangle_B+\lambda_2\left|L G_{l_2}\right\rangle_A|V\rangle_B,
	\label{eq1}
\end{equation}
where $\lambda_i$ represents the normalized complex coefficients,
$|H\rangle$ and $|V\rangle$ correspond to mutually orthogonal horizontal and vertical polarization states.
The polarization states are selected for experimental convenience, where a simple waveplate operation enables conversion between bimeron and skyrmion\cite{banzi}.
Since OAM is directly associated with the photon's spatial mode while polarization correlates with Stokes parameters, the hybrid entangled state shown in equation(\ref{eq1}) can establish a nonlocal vector mapping to form topological structures.
After performing the measurement of photon A at position $\left|\overrightarrow{r_A}\right\rangle$ and neglecting the global phase, the quantum state of photon B can be expressed as:
\begin{equation}
	\left|\Psi_{B \mid A}\right\rangle=\frac{|H\rangle_B+\mu|V\rangle_B}{\sqrt{1+|\mu|^2}},
    \label{eq2}
\end{equation}
with
\begin{equation}
	\mu(r, \phi)=\frac{\lambda_2\left|L G_{l_2}(r)\right| e^{i\left(l_2-l_1\right) \phi}}{\lambda_1\left|L G_{l_1}(r)\right|}=f(r) e^{i \Phi(\phi)},
\end{equation}
where $f(r)=|\mu(r, \phi)|$, $\Phi(\phi)=(l_2-l_1)\phi$.
Obtaining the quantum Stokes parameters is a key step in analyzing the topological properties of quantum Skyrmions. This is achieved by extracting these parameters from the reconstructed two-photon density matrix, using the following computational procedure:
\begin{equation}
	S_j=\left\langle\mid \overrightarrow{r_A}\right\rangle\left\langle\overrightarrow{r_A} \mid \otimes \sigma_{B, j}\right\rangle=\left\langle\Psi_{B \mid A}\right| \sigma_{B, j}\left|\Psi_{B \mid A}\right\rangle,
	\label{eqqst}
\end{equation}
where $\sigma_{B, j}$ denotes the usual Pauli matrices.
By substituting equation (\ref{eq2}) into equation (\ref{eqqst}), the Stokes parameters of the quantum Skyrmion can be obtained as follows:
\begin{equation}
	\begin{aligned}
		& S_x(r)=\frac{2 f(r) \cos \Phi}{1+f(r)^2} \\
		& S_y(r)=\frac{-2 f(r) \sin \Phi}{1+f(r)^2} \\
		& S_z(r)=\frac{1-f(r)^2}{1+f(r)^2}.
	\end{aligned}
	\label{eq5}
\end{equation}
Due to the spatial dependence of the Stokes parameters in the position representation, the calculation of the quantum skyrmion number must be based on the following derivation\cite{skst1}:
\begin{equation}
	N=\frac{1}{4 \pi} \int_S S \cdot\left(\frac{\partial S}{\partial r} \times\left(\frac{1}{r} \frac{\partial S}{\partial \phi}\right)\right) r d r d \phi.
    \label{sknumb}
\end{equation}
By exploiting the properties of Laguerre-Gaussian functions, the explicit expression for the topological charge of the quantum Skyrmion can be derived as follows:
\begin{equation}
	N=\left(l_2-l_1\right)\left(\frac{1}{1+f^2(0)}-\frac{1}{1+f^2(\infty)}\right).
\end{equation}
In the case of $|l_{1}| \neq |l_{2}|$, a nontrivial topological structure is obtained, with the quantum Skyrmion's topological charge being $N=\pm(l_2-l_1)$.

\subsection{Generation of classical optical skyrmions.} We experimentally investigated the robustness of optical Stokes skyrmions in a classical turbulent channel with the setup shown in Figure \ref{fig:Classical results}. Figure \ref{fig:Classical results} \textbf{a} shows an example of the digital phase hologram $H(x,y)$ used to spatially structure the beam using a complex amplitude modulation scheme \cite{arrizon2007pixelated}, 
\begin{equation}
    H(x,y) = J^{-1}_1 (A(x,y)) \sin[\phi(x,y) + 2\pi(G_x x + G_y y) ] \,,
\end{equation}
where $A(x,y)$ and $\phi(x,y)$ is the amplitude and phase of the desired field respectively, $J^{-1}_1$ is the inverse Bessel function of the first kind and $G_{x(y)}$ are the grating frequencies in the horizontal and vertical directions. We encoded a hologram for each of the component scalar modes onto each half of the SLM allowing us to shape each component of the vector beam separately and independently apply and dynamically change the turbulence phase mask on the beam as described in Ref. \cite{peters2025structured}. 

The experimental setup is shown in Figure \ref{fig:Classical results} \textbf{b}, where a horizontally polarized Gaussian beam from a HeNe laser ($\lambda = 633$~nm) was expanded using a $10\times$ objective lens $L_1$ and collimated using a $f=250$~mm lens $L_2$ before impinging onto the screen of a liquid crystal on silicon SLM. The SLM was encoded with holograms for a Gaussian beam of waist $w_0 = 0.5$~mm on the one half and an $LG_l$ beam of variable OAM $l$ and a constant embedded Gaussian beam waist of $w_0=0.5$~mm on the other. The employed complex amplitude modulation scheme generates the desired optical field in the first diffraction order of the digital hologram. This was isolated with the use of a spatial filtering iris and a 4f imagine system consisting of two lenses $L_3$ and $L_4$, both of focal length $f=300$~mm. To create a vectorial superposition, both beams were passed through a HWP to convert them to diagonal polarization and then through a Sagnac interferometer to isolate and coaxially combine the orthogonal polarization components, forming a vector beam of the form $\ket{\Psi} = \left( \ket{LG_0}\ket{V} + \ket{LG_l}\ket{H} \right) / \sqrt{2}$. The encoded skyrmion/wrapping number of the beam is subsequently given by the OAM of the horizontally polarized spatial mode, i.e. $N=l$. 

Figure \ref{fig:Classical results} \textbf{c} shows results for the measured $N$ in the near-field. These were obtained by using a second 4f imaging system consisting of lenses $L_5$ and $L_6$ (both of focal length $f=300$~mm) to image the beam from image plane of the SLM just after the output of the Sagnac interferometer to a polarization sensitive camera capable of measuring all 4 linear polarization projection simultaneously. A quarter-wave plate (QWP) was used to measure the circular polarization projections. These projections allowed for the spatially varying Stokes parameters of the beam to be measured and the skyrmion number was calculated according using the method described in Ref. \cite{mcwilliam2023topological}. The far-field results presented in Figure \ref{fig:Classical results} \textbf{d} were measured with use of  lens $L_7$ of focal length $f=1000$~mm, which Fourier transformed the image plane after the Sagnac interferometer onto the polarisation sensitive camera for full stokes polarimetry measurements. The propagation results shown in Figure \ref{fig:Classical results} \textbf{e} require propagation in two stages. First, the desired modes were numerically propagated using angular spectrum propagation \cite{schmidt2010numerical} half ($0.5$~m) of the desired channel length. The propagated version of the field and the turbulence phase screen were then combined before being used to generate the hologram. The modulated light then propagated through the same 4f imaging system, spatial filter and Sagnac interferometer as before, and then propagated the remaining channel length ($0.5$~m) to the polarisation camera.

\clearpage
\appendix

\setcounter{section}{0}
\setcounter{figure}{0}
\setcounter{table}{0}
\setcounter{equation}{0}
\setcounter{footnote}{0}
\renewcommand{\thesection}{S\arabic{section}}
\renewcommand{\thefigure}{S\arabic{figure}}
\renewcommand{\thetable}{S\arabic{table}}
\renewcommand{\theequation}{S\arabic{equation}}

\section*{Supplementary: Topologically protected quantum skyrmions against turbulence.} 
It can be demonstrated from equation (6) in the main text that the skyrmion number remains invariant under coordinate transformations. 
\begin{figure*}[htb]
\centering\includegraphics[width=\linewidth]{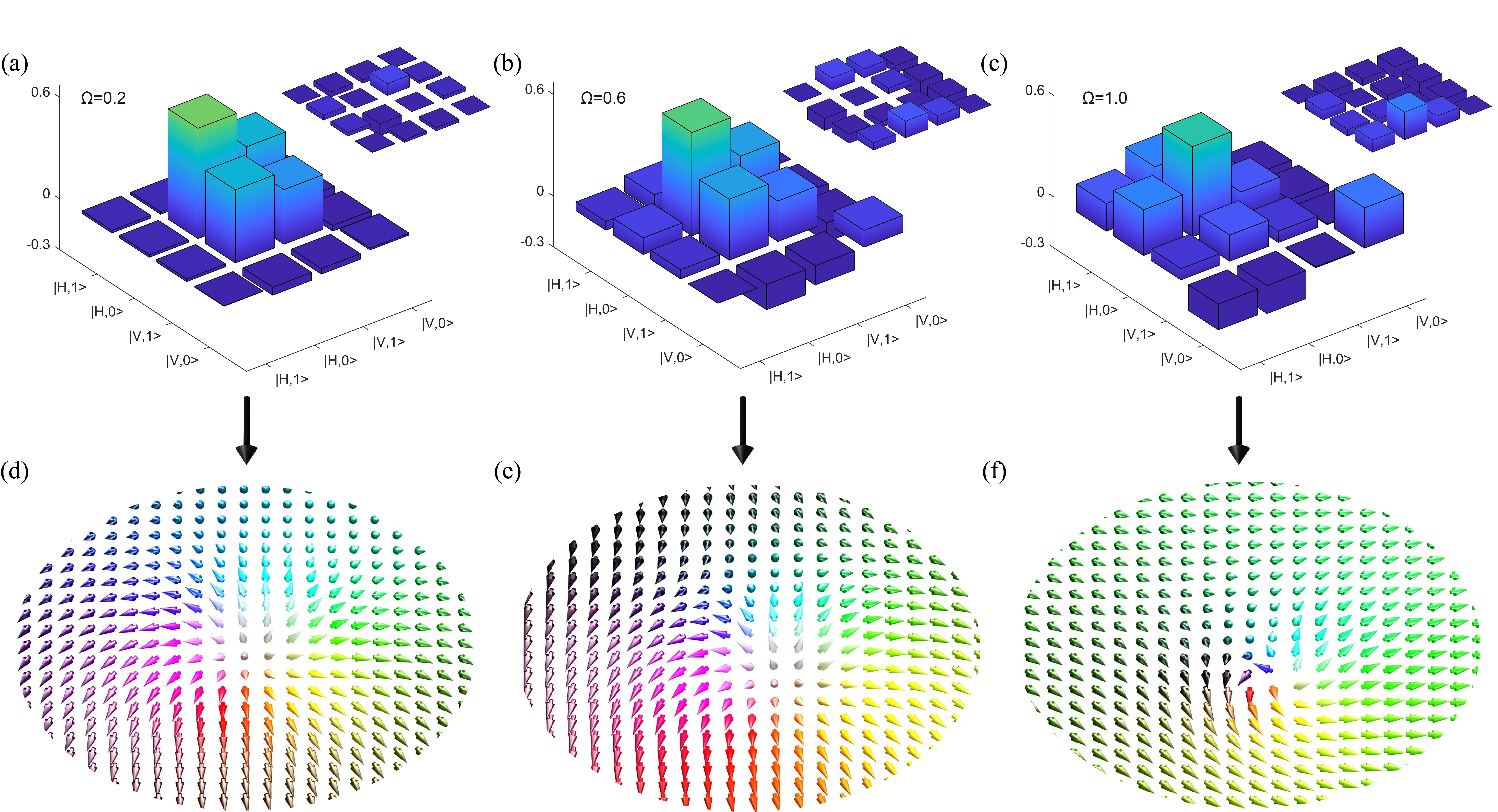}
	\caption{(a)-(c) present the quantum state tomography reconstruction results of skyrmions under varying turbulence intensities, $\Omega=2w/r_0$. (d)-(f) corresponding texture structures.}
    \label{sup1}
\end{figure*}
Therefore, to prove the topological robustness of quantum skyrmions against turbulent disturbances, it suffices to show that the influence of turbulence on quantum states can be expressed in terms of coordinate transformations.

After turbulent evolution, the quantum state can be expressed as:
\begin{equation}
\begin{aligned}
|\Psi\rangle & =\lambda_1\left|L G_{l_1}\right\rangle_A|H\rangle_B+\lambda_2\left|L G_{l_1}\right\rangle_A|V\rangle_B \\
& +\lambda_3\left|L G_{l_2}\right\rangle_A|H\rangle_B+\lambda_4\left|L G_{l_2}\right\rangle_A|V\rangle_B,
\end{aligned}
\end{equation}
where $\lambda_i$ represents the normalized complex coefficients.
The impact of quantum state evolution on subsequent analysis is primarily manifested in:
\begin{equation}
\begin{aligned}
\mu^{\prime}(r, \phi) & =\frac{\lambda_2\left|L G_{l_1}(r)\right|+\lambda_4\left|L G_{l_2}(r)\right| e^{i\left(l_2-l_1\right) \phi}}{\lambda_1\left|L G_{l_1}(r)\right|+\lambda_3\left|L G_{l_2}(r)\right| e^{i\left(l_2-l_1\right) \phi}} \\
& =f^{\prime}(r, \phi) e^{i \Phi^{\prime}(r, \phi)},
\end{aligned}
\label{eq10}
\end{equation}
where $f^{\prime}(r,\phi)=|\mu^{\prime}(r, \phi)|$, $\Phi^{\prime}(r,\phi)=arg(\mu^{\prime}(r, \phi))$.
Given that the turbulent perturbation terms make direct analytical solutions exceptionally challenging, we instead employ a coordinate transformation approach. 
Based on the principle of topological invariance, it suffices to construct a conformal transformation satisfying the following equation to simplify the problem.
\begin{equation}
\mu(r_1, \phi_1)=\mu^{\prime}(r, \phi).
\end{equation}
Considering the mathematical properties of Laguerre-Gaussian beams\cite{guo2023radial}, the following simplification can be adopted:
\begin{equation}
\varepsilon(\mathrm{r})=\frac{\left|\mathrm{LG}_{l_2}\left(\mathrm{r}\right)\right|}{\left|\mathrm{LG}_{l_1}\left(\mathrm{r}\right)\right|}=\sqrt{\frac{\left|l_1\right|!}{\left|l_2\right|!}}\left(\frac{\sqrt{2} \mathrm{r}}{\omega_0}\right)^{\left|l_2\right|-\left|l_1\right|},
\end{equation}
where $\omega_0$ represents the beam waist. The solution can be obtained through coordinate transformation as follows:
\begin{equation}
\begin{aligned}
& r_1=\frac{\omega_0}{\sqrt{2}} \cdot\left[\sqrt{\frac{\left|l_2\right|!}{\left|l_1\right|!}} \cdot\left(\frac{\lambda_1}{\lambda_2} \cdot\left|\frac{\lambda_2+\lambda_4 \varepsilon(r) e^{i\left(l_2-l_1\right) \phi}}{\lambda_1+\lambda_3 \varepsilon(r) e^{i\left(l_2-l_1\right) \phi}}\right|\right)^{\frac{1}{\left|l_2\right|-\left|l_1\right|}}\right] \\
& \phi_1=\frac{1}{l_2-l_1} \arg \left(\frac{\lambda_2+\lambda_4 \varepsilon(r) e^{i\left(l_2-l_1\right) \phi}}{\lambda_1+\lambda_3 \varepsilon(r) e^{i\left(l_2-l_1\right) \phi}}\right).
\end{aligned}
\end{equation}
The topological invariance of the skyrmion number under coordinate transformations fundamentally underlies the intrinsic robustness of quantum skyrmions against turbulent perturbations.

\section*{Supplementary: Quantum State tomography} 
To accurately obtain complete information about the quantum state, we employed a maximum likelihood estimation-based quantum state tomography technique in our experimental research\cite{kaznady2009numerical}. According to the fundamental principles of quantum information theory, any density matrix satisfying the conditions of being Hermitian, non-negative, and unit-trace can be uniquely expressed in the following form through Cholesky decomposition:
\begin{equation}
\hat{\rho}_{\text {ideal }}(\vec{t})=\frac{T(\vec{t})^{\dagger} T(\vec{t})}{\operatorname{Tr}\left\{T(\vec{t})^{\dagger} T(\vec{t})\right\}}.
\end{equation}
For an n-qubit quantum system, the matrix $T(t)$ is a $2^n\times2^n$ matrix characterized by $4^n$ parameters $t = {t_1, t_2, ..., t_{4n}}$. Specifically:
\begin{equation}
T(\vec{t})=\left[\begin{array}{cccc}
t_1 & 0 & 0 & 0 \\
t_{2^n+1}+i t_{2^n+2} & t_2 & 0 & 0 \\
\vdots & & \ddots & \vdots \\
t_{4^n-1}+i t_{4^n} & \cdots & t_{2^{n+1}-4}+i t_{2^{n+1}-3} & t_{2^n}
\end{array}\right].
\end{equation}
The crucial next step involves iteratively optimizing the parameter set $t$ based on experimental data $n_\nu$ obtained from measurement operators $\hat{\Pi}_\nu$, in order to reconstruct the density matrix estimate that most closely approximates the true quantum state.
More precisely, the optimal parameter set $t$ is determined by minimizing the following objective function:
\begin{equation}
\mathcal{L}(\vec{t})=\frac{1}{2} \sum_{\nu=0}^{4^n-1} \frac{\left[ \operatorname{Tr}\left\{\hat{\Pi}_\nu \hat{\rho}_{\text {ideal }}(\vec{t})\right\}-n_\nu\right]^2}{n_\nu}.
\end{equation}
We successfully prepared the quantum skyrmion state in experiments, which can be expressed as:
\begin{equation}
|\Psi\rangle=|0\rangle_A|H\rangle_B+|1\rangle_A|V\rangle_B,
\end{equation}
where $|0\rangle$ and $|1\rangle$ represent the topological charge numbers of OAM, respectively.
The experimentally reconstructed density matrices under varying turbulence intensities are presented in Figure \ref{sup1}(a)-(c), exhibiting distinct structural characteristics that correlate with turbulence strength.
Upon successfully reconstructing the density matrix $\rho$ through quantum state tomography, we obtain complete statistical information about the quantum state. To quantitatively characterize the entanglement properties of the system, we employ the widely-used entanglement measure for biphoton systems – Concurrence, defined as follows:
\begin{equation}
C(\rho)=\max \left\{0, \lambda_1-\lambda_2-\lambda_3-\lambda_4\right\},
\end{equation}
where $\lambda_i$ are the eigenvalues in descending order of the operator $R = \sqrt{\sqrt{\rho}\,\tilde{\rho}\,\sqrt{\rho}}$, with $\tilde{\rho} = (\sigma_y \otimes \sigma_y) \rho^* (\sigma_y \otimes \sigma_y)$, $\sigma_y$ represent the Pauli-Y operator. The concurrence ranges continuously from 0 for separable states to 1 for maximally entangled states.

Based on the reconstructed density matrix, we can further analyze the spin texture characteristics of quantum skyrmions and calculate their topological charge (skyrmion number). The quantum Stokes parameters can be quantitatively analyzed through the following fundamental expressions:\cite{quansk1}:
\begin{equation}
S_j=\operatorname{Tr}\left(|\bar{r}\rangle_A\left\langle\left.\bar{r}\right|_A \otimes \sigma_{B, j} \rho\right)\right.
\end{equation}
The experimentally measured density matrix exhibits skyrmion textures as shown in the Figure \ref{sup1}(d)-(f), demonstrating clear topological characteristics in the spin configuration.
Upon accurately obtaining the quantum Stokes parameters, the skyrmion number of the quantum skyrmion can be determined.

\section*{Supplementary: Simulation of atmospheric turbulence}
In our experimental framework, we model atmospheric turbulence as refractive index fluctuations induced by microscopic temperature and pressure variations. These perturbations initially manifest as phase fluctuations that evolve into compound phase-amplitude modulations during optical propagation. Our implementation employs the thin phase screen approximation, where the turbulence strength ischaracterized by the dimensionless ratio $\Omega = 2w/r_0$, with $w$ representing the second moment beam waist and $r_0$ denoting Fried's atmospheric coherence length parameter, given by \cite{fried1966optical,ndagano2017characterizing}:
\begin{equation}
r_0=0.185 \left(\frac{\lambda}{C_n^2z}\right)^{3/5} ,
\end{equation}
where $C_n^2$ is the refractive index structure constant, $\lambda$ is the wavelength and $z$ is the channel length. The strength of turbulent media can be quantitatively characterized by the Strehl ratio (SR). Under the single phase-screen approximation model, its mathematical expression is given by:
\begin{equation}
\mathrm{SR} \cong \frac{1}{ \left[1+\left(2w / r_0\right)\right]^{5 / 3}}.
\end{equation}

Given the stochastic nature of atmospheric refractive index fluctuations, a statistical description is essential. In this study, we adopt two approaches. Due to its computational efficiency, we made use of a spatial frequency domain approach for the quantum measurements. This approach makes use of fast Fourier transforms (FFTs) to generate random spectral components based on a prescribed power-law spectrum. The methodology centers on implementing the Fourier transform of the refractive index covariance function. In the Fourier domain, the power spectral density of refractive index fluctuations is expressed as:
\begin{equation}
    \Phi_n(\kappa) = 0.0033 C_n^2 k^{-11/3}.
\end{equation}
This formulation represents the renowned Kolmogorov power spectral density model, where $k$ denotes the spatial frequency. 
The power spectral density function provides the fundamental basis for optimally simulating atmospheric turbulence effects through stochastic sampling of spatial frequency components. The implementation involves generating turbulence phase screens via Fourier-domain synthesis, achieved by encoding the Fourier transform of the product between the power spectrum and a complex random field.
\begin{figure*}[htb]
\centering\includegraphics[width=\linewidth]{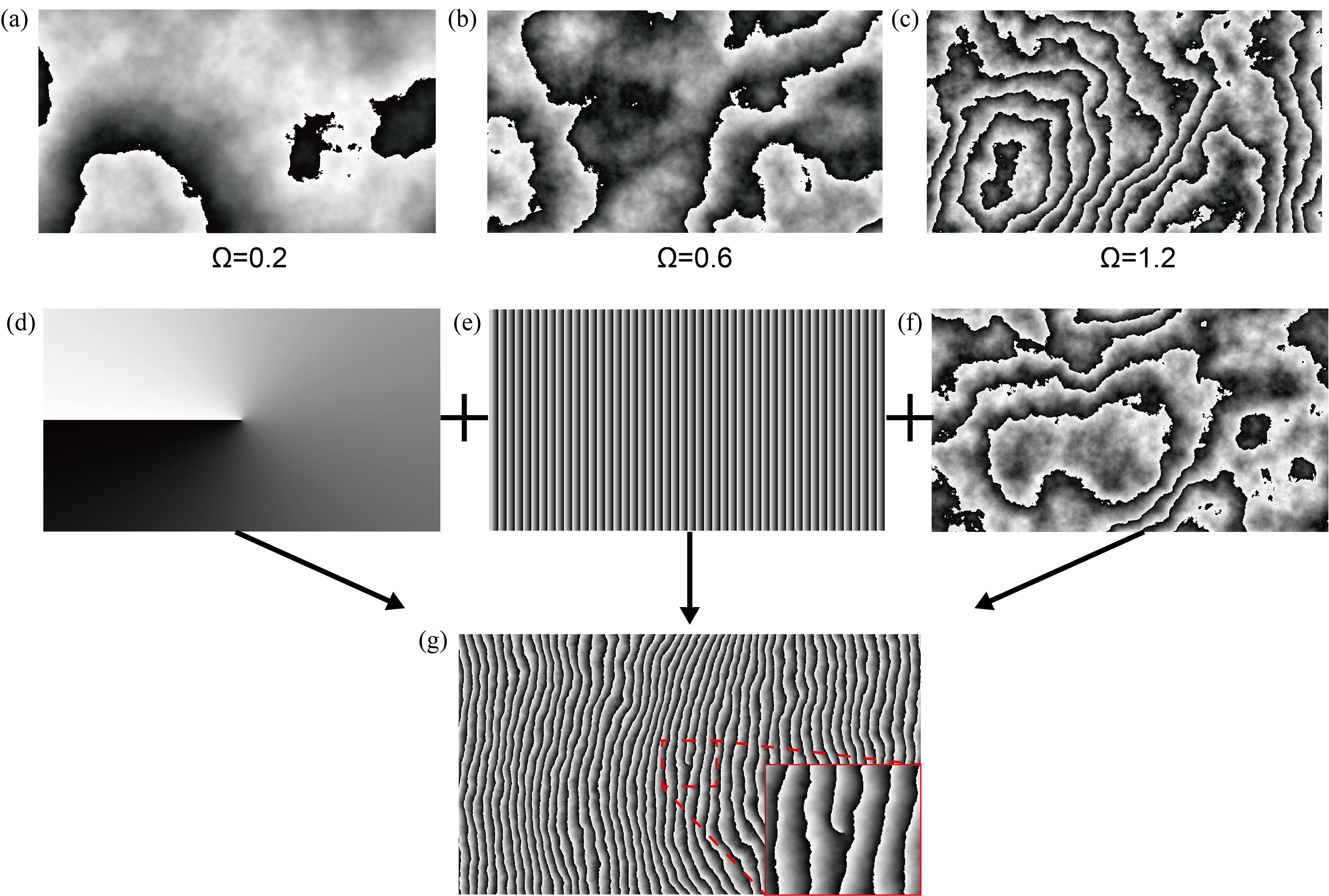}
	\caption{\textbf{Grating configurations.} (a)-(c) respectively display dynamic grating diffraction patterns under three turbulence regimes: weak ($\Omega=0.2$), moderate ($\Omega=0.6$), and strong ($\Omega=1.2$). (d) shows a vortex phase grating with topological charge $l=1$. (e) presents a blazed grating. (f) display grating patterns with ($\Omega=1.0$). (g) depicts the composite grating pattern (1920×1080 pixels) ultimately loaded onto the spatial light modulator.}
    \label{sup2}
\end{figure*}

The phase screens corresponding to different turbulence strengths are presented in Figure \ref{sup2}(a)-(c). In our experimental setup, we employ a spatial light modulator to perform projective measurements on photon A, which necessitates the implementation of a composite grating incorporating three distinct components: (i) vortex phase grating, (ii) blazed grating, and (iii) turbulence-modulated grating, as schematically illustrated in Figure \ref{sup2}(d)-(g).

The classical simulations and experiments made use of Noll phase screens. The spectral domain is popular due to its computational efficiency achieved by leveraging optimized FFT algorithms. However, this approach is known to overestimate the high frequency spatial contributions and underestimate the low frequency components of turbulence \cite{bachmann2025accurate}. This results in high frequency intensity perturbations in the propagation of optical beams leading to higher noise and falsely increases the number of phase and polarization singularities that are detected. In contrast, the Noll phase screens make use of Zernike which form a complete and orthogonal set of the unit disk and are commonly used in simulated and real world adaptive optics applications to correct for the wavefront aberrations induce by atmopsheric turbulence. They are given analytically by,
The analytical form for the Zernike polynomials is therefore given as follows \cite{born2013principles},
\begin{equation}
\begin{aligned}
    &Z_{\text{even \,j}} (r,\theta) &=& \sqrt{n+1} \, R_n^m(r) \sqrt{2} \, \cos{(m\theta)} \\
    &Z_{\text{odd \,j}} (r,\theta) &=& \sqrt{n+1} \, R_n^m(r) \sqrt{2} \, \sin{(m\theta)} \\
    &Z_{\text{j}} (r,\theta) &=& \sqrt{n+1} \, R_0^m(r)\,,
\end{aligned}
\begin{aligned}
&\left.\vphantom{\begin{aligned}
    Z_{\text{even \,j}} (r,\theta) &= \sqrt{n+1} \, R_n^m(r) \sqrt{2} \, \cos{(m\theta)} \\
    Z_{\text{even \,j}} (r,\theta) &= \sqrt{n+1} \, R_n^m(r) \sqrt{2} \, \sin{(m\theta)}
  \end{aligned}}\right\rbrace &m \neq 0\,,\\
&\left.\vphantom{\begin{aligned}
    Z_{\text{j}} (r,\theta) &= \sqrt{n+1} \, R_0^m(r) \,,
  \end{aligned}}\right. &m = 0\,,
\end{aligned}
\label{eq:Zernike polynomial}
\end{equation}
where $r$ is the radial coordinate and $\theta$ is the azimuthal coordinate. The radial component is given by,
\begin{equation}
    R_n^m(r) = \sum_{s = 0}^{(n - m)/2} \frac{(-1)^{s}(n - s)!}{s! [(n + m)/2 - s]! [(n - m)/2 - s]!}
    \label{eq:Zernike radial}
\end{equation}
The polynomials have a known covariance in Kolmogorov turbulence~\cite{noll1976zernike,roddier1990atmospheric}, which can be used to randomly sample weighting coefficients to build phase screens that statistically replicate the effects of atmospheric turbulence. Such an approach is computationally slower but shows better agreement with theoretical predictions of the Kolmogorov model. Full details in the implementation of this method can be found in Ref. \cite{peters2025structured}.

\section*{Supplementary: Calculation of Skyrmion Number from Experimental Data} \label{sec:experimental_determination}

\noindent \textbf{Contour integral calculation of the skyrmion number.} Typical approaches to calculating the wrapping number of given field involve directly computing the surface integral. Due to the severe distortions induced by atmospheric turbulence on the beams' intensity profile, we made used of an alternative approach initially proposed by McWilliam \textit{et al.} \cite{McWilliam2023topological}. This approach involves using a contour integral to reframe the  computation as follows,
\begin{equation}
     N = \frac{1}{2} \left( \sum_{j} S_{z}^{(j)} N_{j} - \bar{S_{Z}^{\infty}} N_{\infty} \right) \,,
     \label{eq:LineInt}
\end{equation}
where $N_j$ is the charge of individual phase singularity at position $j$ in the field $S_x + iS_y $,  $S_z^{(j)}$ is the value of the Stokes parameter $S_z$ at the point $j$, $N_{\infty}$ is the result of the contour integral at infinity and $S_z^{(\infty)}$ is the value of the Stokes parameter $S_z$ as $r\rightarrow \infty$. Any of the Stokes parameters ($S_1$, $S_2$ and $S_3$) can take the place of $S_z$, with the other two ordered taking the place of $S_x$ and $S_y$.\\

\noindent \textbf{Classical and quantum stokes parameters.} In order to make use of Equation \ref{eq:LineInt}, the Stokes parameters must be determined from experimental measurements. In the case of quantum skyrmions, the Stokes parameters are simply the observables of the Pauli matrices which can be calculate from the experimentally reconstructed density matrix $\sigma_{B, j}$ as follows,.
\begin{equation}
	S_j=\left\langle\mid \overrightarrow{r_A}\right\rangle\left\langle\overrightarrow{r_A} \mid \otimes \sigma_{B, j}\right\rangle=\left\langle\Psi_{B \mid A}\right| \sigma_{B, j}\left|\Psi_{B \mid A}\right\rangle.
	\label{eqqst}
\end{equation}
Classical Stokes parameters are obtained through traditional Stokes polarimetry using six experimentally measured polarization intensity projections,
\begin{eqnarray} \label{eq:stokesS0}
    s_{0} &=& I_{H} + I_{V} \\
\label{eq:stokesS1} 
    s_{1} &=& I_{H} - I_{V}\\
\label{eq:stokesS2}
    s_{2} &=& I_{D} - I_{A}\\
\label{eq:stokesS3}
    s_{3} &=& I_{R} - I_{L} \,.
\end{eqnarray}
The subscripts H, V, D, A, R and L represent horizontal, vertical, diagonal, antidiagonal, right circular and left circular polarizations respectively. A polarization sensitive camera was able to measure the four linear polarization intensity projections. A quarter-wave plate was placed in from the of the camera in order to measure the circular intensity projections \cite{cox2023real}. Equation \ref{eq:LineInt} requires the locally normalized Stokes parameters $S_j$ which were computed from the experimentally obtained Stokes parameters $s_j$ according to,
\begin{equation}
\label{eq:localNormStokes}
    S_j =\frac{s_j}{\sqrt{s_1^2+s_2^2+s_3^2}}
\end{equation}
\\

\noindent \textbf{Polarization singularities.} The contour integral can be taken over one of three polarization fields,
\begin{eqnarray} \label{eq:polphase1}
P_{1} &=& S_{2} + iS_{3}\\
\label{eq:polphase2}
P_{2} &=& S_{3} + iS_{1}\\
\label{eq:polphase3}
P_{3} &=& S_{1} + iS_{2}\,.
\end{eqnarray}
While all three are theoretically equivalent, numerical and practical considerations may cause one choice to perform far more reliably than others. In this work. $P_1$ exhibited erratic behaviour over various realizations as compared to $P_2$ and $P_3$ and so was excluded. Once the polarization field were obtained, the positions and charges of the phase singularities of these fields were determined using a numerical equivalent to the curl $\nabla \times$ operation (first proposed in Ref \cite{chen2007detection}) termed the circulation $D$. The circulation is defined as,
\begin{eqnarray}
    D^{m,n} = &\frac{d}{2}& ( G_x^{m,n} + G_x^{m,n+1} +G_y^{m,n+1} + G_y^{m+1,n+1} \nonumber \\ &-  &G_x^{m+1,n+1} - G_x^{m+1,n} -G_y^{m+1,n} \nonumber \\
    &-& G_y^{m,n})\,.
\end{eqnarray} 
Here, $D^{m,n}$ represents the value of the circulation of the pixel in the $n$-th row and $m$-th column. $G_x^{m,n}$ and $G_y^{m,n}$ are the phase gradient in the horizontal and vertical direction of the pixel in the $n$-th row and $m$-th column, respectively and $d$ is the pixel size. Typically, the circulation will return a 0 value if there is no singularity at that pixel and a nonzero value if there is. The magnitude of the circulation indicates the charge of the singularity and the sign indicates the direction/handedness of the singularity. These values were then substituted into Equation \ref{eq:LineInt} to calculate N.\\

\section*{Supplementary: Post-processing of classical experimental data}

 The local normalization of the Stokes parameters in Equation \ref{eq:localNormStokes} is necessary to ensure the accurate calculation of the wrapping number. However, it also results in the amplification of random noise in low intensity regions of experimental data. This noise can be caused by ambient environmental light and the shot noise of the detector. If it is not  considered or filtered off, it will artificially increase the number of polarization states present the measurement and result in an inaccurate determination of the skyrmion number. We therefore implemented a standard post-processing procedure following the one outlined in the supplementary of  Ref. \cite{peters2025seeing} that was calibrated on unaberrated beams for each of the configurations tested: near-field, far-field and single phase screen propagation. The parameters in each case were kept constant irregardless of the turbulence strength or incident $N$, ensuring the measurement system and procedure was agnostic of the specific topology or turbulence strength.

 For the near-field results, an intensity based threshold was implemented. The singularities of the polarization field was calculated, and any singularity found in regions where the measured intensity was below the noise floor were disregarded.  To determine the noise floor, regions near the edges of the captured CCD images, far removed from the generated beam, were isolated and the intensity values averaged. The average noise value over multiple measurements was found to be $\approx 3\%$ of the maximum intensity. Therefore, any singularity detected in regions with intensities lower than 3\% of the maximum beam signal were disregarded from the calculation of the skyrmion number. This was kept constant for all of the near-field measurements. The far-field measurements implemented the same intensity based threshold and also included a low-pass 2D Gaussian filter with kernel of standard deviation $\sigma = 1220$~m$^{-1}$. The far-field saw the manifestation of the high frequency intensity features due to the propagation of the light. These high frequency intensity features result in additional singularities which affect the wrapping number calculation, thus necessitating the need for the Gaussian filter. The kernel size and threshold were kept constant for all of the far-field measurements. The propagation results made use of an intensity threshold of 3\% and Gaussian filter with kernel $\sigma = 732$~m$^{-1}$ which again was kept constant for all of the propagation measurements.

 \section*{Supplementary: Parameters for multiple phase screen simulations}

The multiple phase screen simulations were performed in MATLAB. We made use of an initial grid size of $1024\times1024$ pixels with pixel size $\delta x=160~\mu$m. The maximum radius of the Zernike polynomials was set to $R_{max}=82$~mm and 128 Zernike modes were used in the construction of the turbulence phase screens. The skyrmions were generated through vectorial combinations of LG beams, with the embedded Gaussian beam waist $w_0 = 25$~mm. The length of the channel was $L = 1$~km with a spacing of $\Delta z=200$~m between the five screens. Channels with Rytov variances $\sigma_R^2 = 1.0,\,1.5,\,2.0,\,2.5$ and $3.0$ were tested, with corresponding Fried parameters for each unit cell being $r_{0,s}=50,39,33,29$ and $26$~mm respectively and thus total channel Fried parameters of $r_0 = 19.0,\,14.9,\,12.6,\,11.0$ and $9.85$~mm respectively. Each simulation saw 100 independent, random realizations for each of the five phase screens.

\end{document}